\documentclass[preprint,3p,12pt]{elsarticle}

\usepackage{lineno,hyperref}
\modulolinenumbers[5]

\usepackage{hyperref}

\usepackage{epstopdf}

\usepackage{mathrsfs}
\usepackage{amssymb}
\usepackage{amsfonts}
\usepackage{latexsym}
\usepackage{amsmath}
\usepackage{xcolor}
\usepackage{wrapfig}
\usepackage{floatflt}

\usepackage{graphicx}

\def\lb{\label}

\newcommand{\er}[1]{\textrm{(\ref{#1})}}

\newtheorem{theorem}{\bf Theorem}[section]
\newtheorem{lemma}[theorem]{\bf Lemma}
\newtheorem{definition}[theorem]{\bf Definition}

\def\a{\alpha}  \def\cA{{\mathcal A}}       
\def\b{\beta}   \def\cB{{\mathcal B}}       
  \def\cC{{\mathcal C}}       \def\mC{{\mathscr C}}

         \def\mF{{\mathscr F}}
           \def\mG{{\mathscr G}}
          \def\mH{{\mathscr H}}
    \def\cI{{\mathcal I}}

\def\l{\lambda}        
        
     \def\cO{{\mathcal O}}

\def\s{\sigma}  \def\cR{{\mathcal R}}       
         \def\mS{{\mathscr S}}
\def\t{\tau}

\def\O{\Omega}

\def\ve{\varepsilon}           

       \def\C{{\mathbb C}}    
    \def\N{{\mathbb N}}   

\def\lt{\biggl}                  \def\rt{\biggr}
\def\ol{\overline}               \def\wt{\widetilde}

\let\ge\geqslant                 \let\le\leqslant

\def\iy{\infty}
               
\def\ss{\subset}                 \def\ts{\times}
\def\pa{\partial}                \def\os{\oplus}
                 \def\ev{\equiv}
        
\def\el2{\ell^{\,2}}             \def\1{1\!\!1}

\def\wh{\widehat}

\def\det{\mathop{\mathrm{det}}\nolimits}

\def\dim{\mathop{\mathrm{dim}}\nolimits}

\def\Tr{\mathop{\mathrm{Tr}}\nolimits}
\def\BBox{\hspace{1mm}\vrule height6pt width5.5pt depth0pt \hspace{6pt}}

\let\ge\geqslant
\let\le\leqslant

\newcommand{\ca}{\begin{cases}}
\newcommand{\ac}{\end{cases}}
\newcommand{\ma}{\begin{pmatrix}}
\newcommand{\am}{\end{pmatrix}}
\def\eq{\begin{equation}}
\def\qe{\end{equation}}
\def\[{\begin{equation}}
\def\]{\end{equation}}

\def\BBox{\hspace{1mm}\vrule height6pt width5.5pt depth0pt \hspace{6pt}}

\begin{document}

\begin{frontmatter}

\title{Determinants and traces in the algebra of multidimensional discrete periodic operators with defects}
\date{\today}

\author
{Anton A. Kutsenko}

\address
{GeoRessources, UMR 7359, ENSG, Vandoeuvre-l\`es-Nancy, 54518,
France; email: akucenko@gmail.com}

\begin{abstract}
As it is shown in previous works, discrete periodic operators with
defects are unitarily equivalent to the operators of the form
$$
 \cA{\bf u}={\bf A}_0{\bf u}+{\bf A}_1\int_0^1dk_1{\bf B}_1{\bf
 u}+...+{\bf A}_N\int_0^1dk_1...\int_0^1dk_N{\bf B}_N{\bf u},\ \ {\bf
 u}\in L^2([0,1]^N,\C^M),
$$
where $({\bf A},{\bf B})(k_1,...,k_N)$ are continuous matrix-valued
functions of appropriate sizes. All such operators form a non-closed
algebra $\mH_{N,M}$. In this article we show that there exist a
trace $\pmb{\t}$ and a determinant $\pmb{\pi}$ defined for operators
from $\mH_{N,M}$ with the properties
$$
 \pmb{\t}(\a\cA+\b\cB)=\a\pmb{\t}(\cA)+\b\pmb{\t}(\cA),\ \
 \pmb{\t}(\cA\cB)=\pmb{\t}(\cB\cA),\ \
 \pmb{\pi}(\cA\cB)=\pmb{\pi}(\cA)\pmb{\pi}(\cB),\ \
 \pmb{\pi}(e^{\cA})=e^{\pmb{\t}(\cA)}.
$$
The mappings $\pmb{\pi}$, $\pmb{\t}$ are vector-valued functions.
While $\pmb{\pi}$ has a complex structure, $\pmb{\t}$ is simple
$$
 \pmb{\t}(\cA)=\lt(\Tr{\bf A}_0,\int_0^1dk_1\Tr{\bf B}_1{\bf A}_1,...,\int_0^1dk_1...\int_0^1dk_N\Tr{\bf B}_N{\bf
 A}_N\rt).
$$
There exists the strong norm under which the closure
$\ol{\mH}_{N,M}$ is a Banach algebra, and $\pmb{\pi}$, $\pmb{\t}$
are continuous (analytic) mappings. This algebra contains
simultaneously all operators of multiplication by matrix-valued
functions and all operators from the trace class. Thus, it
generalizes the other algebras for which determinants and traces was
previously defined.
\end{abstract}


\end{frontmatter}


\section{Introduction}\lb{S0}

Periodic operators with defects play important role in physics and
mechanics of waves, see, e.g., discussions in \cite{K2}. It is shown
in \cite{K3} that these operators are unitarily equivalent to some
multidimensional integral operators which form a non-closed algebra.
In the current paper we try to construct traces and determinants in
this algebra, and we try to find some norms under which these
mappings are continuous.

Traces and determinants of square matrices are familiar to us from
school. The theory of traces and determinants of some classes of
operators acting on infinite dimensional Banach spaces is presented
perfectly in the book \cite{GGK}. Traces and determinants play
important role in various fields. They can be used for determining
the spectrum (zeroes of determinants) and for deriving various trace
formulas, see, e.g.  \cite{BS}, \cite{O}. There is also a general
mathematical interest, see, e.g., \cite{GKZ}, \cite{S}. Usually, the
discussed determinants are scalars because the spectrum of
corresponding operators is discrete. In our case we have the
operators with discrete and continuous spectral components. This
fact leads to vector-valued functional traces and determinants. To
define the determinant we factorize the group of invertible elements
of our algebra into the product of "elementary" subgroups. For each
of the subgroup we determine the scalar functional determinant. The
vector consisting of all such determinants is the final determinant
that we are looking for. The derivative of this determinant at the
identity element is exactly the trace. After that we find a norm
under which the trace (and hence the determinant) is continuous. Our
algebra equipped with this norm becomes a Banach algebra. Let us
start with the definition of the space $L^2_{N,M}$ and the integral
operators $\langle\cdot\rangle_j$:
\begin{definition}\lb{D1a}
Let $L^2_{N,M}:=L^2([0,1]^N,\C^M)$ be the Hilbert space of all
vector-valued (if $M>1$) square-integrable functions ${\bf f}({\bf
k})$ with ${\bf k}=(k_1,...,k_N)\in[0,1]^N$. Define
\[\lb{001}
\langle\cdot\rangle_{j}:=\int_{[0,1]^j}\cdot dk_1...dk_j,\ \ j\le N.
\]
\end{definition}

The algebra of multidimensional periodic operators with defects was
introduced in \cite{K3} as

\begin{definition}\lb{D2a}
The algebra of periodic operators with parallel defects
$$
 \mH_{N,M}={\rm Alg}(\{{\bf A}\cdot\},\langle\cdot\rangle_1,...,\langle\cdot\rangle_N)
$$
is a minimal non-closed subalgebra of the algebra of continuous
linear operators acting on $L^2_{N,M}$, which contains all operators
of multiplication by $M\ts M$ continuous matrix-valued functions
${\bf A}\cdot$ and all integral operators $\langle\cdot\rangle_j$.
\end{definition}

Usually we will omit indices $N,M$, i.e. we will write
$\mH:=\mH_{N,M}$, $L^2:=L^2_{N,M}$. The next theorem proved in
\cite{K3} give simple representation of the operators from $\mH$.

\begin{theorem}\lb{P1}
Each operator $\cA\in\mH$ has a following representation
\[\lb{002}
 \cA {\bf u}={\bf A}_0{\bf u}+{\bf A}_1\langle{\bf B}_1{\bf
 u}\rangle_{1}+...+{\bf A}_N\langle{\bf B}_N{\bf u}\rangle_{N},\ \ {\bf
 u}\in L^2,
\]
where ${\bf A}$, ${\bf B}$ are continuous matrix-valued functions on
$[0,1]^N$ of sizes
\[\lb{003}
 \dim({\bf A}_0)=M\ts M,\ \ \dim({\bf B}_j)=M_j\ts M,\ \ \dim({\bf
 A}_j)=M\ts M_j,\ \ j\ge1
\]
with some positive integers $M_j$. The set of all operators of the
form \er{002} coincides with $\mH$.
\end{theorem}

For convenience, we often will replace the argument ${\bf u}$ with
$\cdot$ in formulas like \er{002}. For example, it can be proved
that the Hermitian adjoint to $\cA$ \er{002} is
\[\lb{0032}
 \cA^*={\bf A}_0^*\cdot+{\bf
 B}_1^*\langle{\bf A}_1^*\cdot\rangle_1+...+{\bf B}_N^*\langle{\bf
 A}_N^*\cdot\rangle_N.
\]

The main question is to find the explicit procedure that can tell us
$\cA\in\mH$ is invertible or non-invertible. One of such procedures
is constructed in \cite{K3}. The inverse operator is also
constructed. It was shown that if $\cA$ is invertible then
$\cA^{-1}\in\mH$. In the current paper, we provide a little modified
version of the procedure from \cite{K3}:

\begin{theorem}\lb{T1}
Let $\cA$ be of the form \er{002}. Then

\underline{Step 0.} Define
\[\lb{005a}
 \pi_0=\det{\bf E}_0,\ \ {\bf E}_0={\bf A}_0.
\]
If $\pi_0({\bf k}^0)=0$ for some ${\bf k}^0\in[0,1]^N$ then $\cA$ is
non-invertible else define
\[\lb{005}
 {\bf A}_{j0}={\bf A}_0^{-1}{\bf A}_j,\ \ j=1,...,N.
\]

\underline{Step 1.} Define
\[\lb{006a}
 \pi_1=\det{\bf E}_1,\ \ {\bf E}_1={\bf I}+\langle{\bf B}_1{\bf A}_{10}\rangle_1.
\]
If $\pi_1({\bf k}_1^0)=0$ for some ${\bf k}_1^0\in[0,1]^{N-1}$ then
$\cA$ is non-invertible else define
\[\lb{006}
 {\bf A}_{j1}={\bf A}_{j0}-{\bf A}_{10}{\bf
 E}_1^{-1}\langle{\bf B}_1{\bf A}_{j0}\rangle_1,\ \ j=2,...,N.
\]

\underline{Step 2.} Define
\[\lb{007a}
 \pi_2=\det{\bf E}_2,\ \ {\bf E}_2={\bf I}+\langle{\bf B}_2{\bf A}_{21}\rangle_2.
\]
If $\pi_2({\bf k}_2^0)=0$ for some ${\bf k}_2^0\in[0,1]^{N-2}$ then
$\cA$ is non-invertible else define
\[\lb{007}
 {\bf A}_{j2}={\bf A}_{j1}-{\bf A}_{21}{\bf
 E}_2^{-1}\langle{\bf B}_2{\bf A}_{j1}\rangle_2,\ \ j=3,...,N.
\]

*********

\underline{Step N.} Define
\[\lb{008a}
 \pi_N=\det{\bf E}_N,\ \ {\bf E}_N={\bf I}+\langle{\bf B}_N{\bf A}_{N,N-1}\rangle_N.
\]
If $\pi_N=0$  then $\cA$ is non-invertible else $\cA$ is invertible.

\end{theorem}

This Theorem can be used for determining the spectrum of the
operator $\cA$. Taking $\pi_{N+1}=0$ and $\l{\bf I}-{\bf A}_0$
instead of ${\bf A}_0$ in the scheme \er{005a}-\er{008a} (or, more
general, ${\bf A}_j(\l,{\bf k})$ instead of ${\bf A}_j({\bf k})$ for
all $j$, see corresponding generalized spectral problems in
\cite{K1}) we can define the function
\[\lb{mult}
 D(\l)=\min\{j: \pi_j=0\ for\ some\ {\bf
 k}_j\in[0,1]^{N-j}\}.
\]
Then the spectrum of $\cA$ is the following set
\[\lb{spec}
 \s(\cA)=\{\l:\ D(\l)\le N\}.
\]
Moreover, the function $D$ shows the "degree" of the spectral point.
For example, if $D(\l)<N$ then $\l$ belongs to the essential
spectrum (in our case this is a continuous part or an eigenvalue of
infinite multiplicity), or if $D(\l)=N$ (the maximum value within
the spectrum) then $\l$ is a point of discrete spectrum. If
$D(\l)=N+1$ (the maximum value) then $\l$ does not belong to the
spectrum.

{\bf Remark on the Floquet-Bloch dispersion curves.} Almost all
papers devoted to the wave propagation through periodic media study
the so-called Floquet-Bloch dispersion curves, see, e.g.,
discussions in \cite{B}, \cite{Kuchment}, and \cite{KKSP1}. These
curves usually describe the dependence of the spectral parameter
(e.g. frequency) of the wave-number ${\bf k}$. For the pure periodic
media this dependence is well-known and can be expressed as
$\l=\l_0({\bf k})$ where $\l_0$ is an implicit function satisfying
$\pi_0\ev\pi_0(\l,{\bf k})=0$. The corresponding waves are called
"propagative" because they have no attenuation. In our case we have
a lot of defects of different dimensions. So there are the waves
which propagate along the defects and exponentially decrease in
perpendicular directions. Depending on the dimension of defects and
of the context this type of waves is usually called "guided",
"surface", "local", "defect", Rayleigh waves, Love waves and so on.
Our method allows us to obtain dispersion equations for such waves.
These are $\l=\l_j({\bf k}_j)$, ${\bf k}_j\in[0,1]^{N-j}$, where
$\l_j$ are implicit functions satisfying $\pi_j\ev\pi_j(\l,{\bf
k}_j)=0$. Note that while $\pi_0$ is a polynomial of $\l$ the other
functions $\pi_j$ are much more complex.

The proof of this theorem gives us an explicit representation of
inverse operator:

\begin{theorem}\lb{C1}
Let $\cA$ be of the form \er{002}. If $\cA$ is invertible then
\[\lb{013}
 \cA=({\bf A}_0\cdot)\circ(\cI+{\bf A}_{1,0}\langle{\bf
 B}_1\cdot\rangle_1)\circ...\circ(\cI+{\bf A}_{N,N-1}\langle{\bf
 B}_N\cdot\rangle_N),
\]
where ${\bf A}_{j,j-1}$ are defined in the scheme
\er{005a}-\er{008a} and $\cI$ is the identity operator. Moreover,
the inverse operator is
\[\lb{015}
 \cA^{-1}=(\cI-{\bf A}_{N,N-1}{\bf E}^{-1}_N\langle{\bf
 B}_N\cdot\rangle_N)\circ...\circ(\cI-{\bf A}_{1,0}{\bf E}^{-1}_1\langle{\bf
 B}_1\cdot\rangle_1)\circ({\bf A}_0^{-1}\cdot).
\]
(We will use $\circ$ to denote the multiplication (composition) of
operators.)
\end{theorem}

Define the following subsets of operators from $\mH$:
\[\lb{010}
 \mH_0=\{{\bf A}\cdot\},\ \ \mH_{j}=\{{\bf A}\langle{\bf B}\cdot\rangle_j\},\ \
 j=1,...,N,
\]
\[\lb{012a}
 \mF_0={\rm Inv}(\mH_0),\ \ \mF_j={\rm Inv}(\{\cI+\cA:\ \cA\in\mH_j\}),\ \ j=1,...,N.
\]
where ${\bf A}$, ${\bf B}$ denote all possible continuous
matrix-valued functions of appropriate sizes (see \er{003}) and
${\rm Inv}$ means all invertible elements of the set. Let us
consider some of their properties:

\begin{theorem}\lb{TG}
The sets $\mF_j$ are groups ($\circ$ is a multiplication) and
\[\lb{012b}
 \mF_0=\{{\bf A}\cdot:\ \ \det{\bf A}\ne0\},\ \ \mF_j=\{\cI+{\bf A}\langle{\bf B}\cdot\rangle_j:\ \
 \det({\bf I}+\langle{\bf B}{\bf A}\rangle_j)\ne0\},
\]
where ${\bf I}$ is the identity matrix of appropriate size. Because
$\det...$ is a function, the expression $\ne0$ assumes everywhere
(for any value of the argument). The inverse operator has the form
\[\lb{012c}
 (\cI+{\bf A}\langle{\bf B}\cdot\rangle_j)^{-1}=\cI-{\bf A}({\bf I}+\langle{\bf B}{\bf
 A}\rangle_j)^{-1}\langle{\bf B}\cdot\rangle_j.
\]
\end{theorem}

The sets \er{010}, \er{012a} are the building blocks for $\mH$:

\begin{theorem}\lb{T3}
The following identities are fulfilled:
\[\lb{alg_pr}
 \mH=\sum_{j=0}^N\mH_j,\ \ {\rm
Inv}(\mH)=\prod_{j=0}^N\mF_j,
\]
where the order of elements in the product is not important.
Moreover, for any $\cA\in\mH$, $\cB\in{\rm Inv}(\mH)$, and any
permutation $(j_0,...,j_N)$ of the set $(0,...,N)$ there exist
unique representations
\[\lb{012}
 \cA=\sum_{j=0}^N\cA_j,\ \cA_j\in\mH_j,\ \ \cB=\cB_{j_0}\circ...\circ\cB_{j_N},\ \cB_{j_N}\in\mF_{j_N}.
\]
\end{theorem}
For given operator $\cA$ of the form \er{002} we have that
$\cA_0={\bf A}_0\cdot$, $\cA_j={\bf A}_j\langle{\bf
B}_j\cdot\rangle_j$ in \er{012}. Also, the representations \er{013},
\er{015} are unique for given order of indices, see \er{012}. To
formulate our main result let us introduce the commutative algebras
of continuous scalar functions
\[\lb{contf}
 \cC_j=\{f:[0,1]^{N-j}\to\C\},\ j=0,...,N-1;\ \ \cC_N=\C;\ \
 \cC=\cC_0\ts\cC_1\ts...\ts\cC_N.
\]

\begin{theorem}\lb{T2}
The mapping (see definitions of $\pi_j$ in \er{005a}-\er{008a})
\[\lb{009}
 \pmb{\pi}=(\pi_0,...,\pi_N):{\rm Inv}(\mH)\to{\rm Inv}(\mC)
\]
is a group homomorphism. Moreover,
$\pmb{\pi}|_{\mF_j}=(1,...,\pi_j,...,1)$.
\end{theorem}
The result \er{009} of this theorem shows us that $\pmb{\pi}$ is an
analogue of the standard determinant of matrices (or matrix-valued
functions). The set ${\rm Inv}(\mC)$ has a simple form, it consists
of non-zero continuous functions. For the one dimensional case the
theory of Fredholm determinants of \{identity $+$ compact
operators\} is well developed, see, e.g., \cite{GGK}. In our case
the situation is complicated by the fact that our perturbations are
not compact in the usual sense. That is why our construction leads
to the vector-valued functional determinant $\pmb{\pi}=(\pi_j)$.
Note that by using the different combinations of $\pi_j$ we can
construct other homomorphisms such as the product $\pi_0...\pi_N$
but they contain less information than $\pmb{\pi}$. The determinant
$\pmb{\pi}(\cA)$ completely describes the spectrum of the operator
$\cA$. For example, we can define the isospectral set of operators
as
\[\lb{iso}
 {\rm Iso}(\cA)=\{\cB:\ \pmb{\pi}(\l\cI-\cA)=\pmb{\pi}(\l\cI-\cB)\ for\ large\
 \l\}.
\]

Along with the vector-valued determinant $\pmb{\pi}$ of invertible
operators of the form \er{002} it is possible to define the
vector-valued trace $\pmb{\t}$ for all operators (invertible and
non-invertible) of the form \er{002}:
\[\lb{016}
 \pmb{\t}(\cA):=\frac{\pa\pmb{\pi}(\cI+t\cA)}{\pa t}|_{t=0}=
 \lim_{t\to0}\frac{\pmb{\pi}(\cI+t\cA)-\pmb{\pi}(\cI)}{t}.
\]
Due to the fact that $\pmb{\pi}$ is a homomorphism the derivative at
other points $\cA\in\mG$ can be found as
\[\lb{017}
 \frac{\pa\pmb{\pi}(\cA+t\cB)}{\pa
 t}|_{t=0}=\pmb{\pi}(\cA)\pmb{\t}(\cA^{-1}\cB),
\]
where the product of vectors means component-wise product. The next
Theorem gives us the explicit formula for $\pmb{\t}$ and provides
its properties.
\begin{theorem}\lb{T4}
For any operator $\cA$ of the form \er{002} the following identity
is fulfilled
\[\lb{018}
 \pmb{\t}(\cA)=(\Tr{\bf A}_0,\langle\Tr{\bf B}_1{\bf A}_1\rangle_1,...,\langle\Tr{\bf B}_N{\bf
 A}_N\rangle_N),
\]
where $\Tr$ means the standard trace of square matrices. Moreover,
the following properties are fulfilled also
\[\lb{019}
 \pmb{\t}(\a\cA+\b\cB)=\a\pmb{\t}(\cA)+\b\pmb{\t}(\cB),\ \ \
 \pmb{\t}(\cA\circ\cB)=\pmb{\t}(\cB\circ\cA)
\]
for any $\cA,\cB$ of the form \er{002} and any $\a,\b\in\C$.
\end{theorem}

Roughly speaking, it can be shown that \er{018} is in good agreement
with the known trace of finite rank operators. In this sense, the
definition of $\pmb{\t}$ (and hence $\pmb{\pi}$) is unique up to
elementary combinations of its components. Let us discuss some trace
norm under which $\pmb{\pi}$ and $\pmb{\t}$ are continuous and a
completion of $\mH$ is a Banach algebra. For an operator $\cA$ of
the form \er{002} define the functions
\[\lb{022}
 g_j({\bf k}_j)=\sum_{n=1}^{M_j}\sqrt{\l_{nj}({\bf k}_j)},\ \ {\bf
 k}_j=(k_{j+1},...,k_N)\in[0,1]^{N-j},\ \ j=0,...,N
\]
(for $j=N$ there is no dependence on ${\bf k}_N$ and $f_N$ is just a
number), where $\{\l_{nj}\}_{n=1}^{M_j}$ are eigenvalues of $M_j\ts
M_j$ matrices ${\bf C}_j$ defined by
\[\lb{023}
 {\bf C}_0:={\bf A}_0^*{\bf A}_0,\ \ {\bf C}_j:=\langle{\bf B}_j{\bf
 B}_j^*\rangle_j\langle{\bf A}_j^*{\bf A}_j\rangle_j.
\]
All $\l_{nj}$ are non-negative because they are singular values of
the operator ${\bf A}_j\langle{\bf B}_j\cdot\rangle_j$. Define the
following non-negative function
\[\lb{024}
 \|\cA\|_{\rm tr}=\max_{{\bf k}_0\in[0,1]^N}g_0({\bf k}_0)+\max_{{\bf k}_1\in[0,1]^{N-1}}g_1({\bf
 k}_1)+...+g_N.
\]
We also denote
\[\lb{025a}
 \|{\bf
 f}\|_{\rm c}=\max_j\max_{{\bf k}_j\in[0,1]^{N-j}}|f_j({\bf k}_j)|,\ \
 {\bf f}=(f_0,...,f_N)\in\mC,
\]
where $\mC$ is a commutative Banach algebra defined in \er{contf}
with an element-wise multiplication.

\begin{theorem}\lb{T5}
The function $\|\cdot\|_{\rm tr}$ is a norm on $\mH$. The
corresponding completion $\ol{\mH}$ is a Banach algebra with
\[\lb{025}
 \|\cA\circ\cB\|_{\rm tr}\le\|\cA\|_{\rm tr}\|\cB\|_{\rm tr},\ \
 \|\cA\|\le\|\cA\|_{\rm tr},\ \ \forall\cA,\cB\in\ol{\mH},
\]
where $\|\cdot\|$ denotes the standard operator norm. The mappings
$\pmb{\t}$ and $\pmb{\pi}$ are continuous and have continuous
extensions
\[\lb{026}
 \pmb{\t}:\ol{\mH}\to\mC,\ \ \pmb{\pi}:{\rm Inv}(\ol{\mH})\to{\rm Inv}(\mC).
\]
The norm of $\pmb{\t}$ (as a linear operator) is $1$ and
$\forall\cA\in\ol{\mH}$ we have
\[\lb{020}
 \ln\pmb{\pi}(\l\cI-\cA)=(\ln
 \l)\pmb{\t}(\cI)-\sum_{n=1}^{\iy}\frac{\pmb{\t}(\cA^n)}{n\l^n},\ \ |\l|>\|\cA\|_{\rm tr},
\]
where $\ln$ in the left-hand side means element-wise logarithm and
$\pmb{\t}(\cI)=(M,0,...,0)$.
\end{theorem}
Note that \er{020} with \er{018} allow us to obtain more convenient
representation of the set \er{iso}
\[\lb{021}
 {\rm Iso}(\cA)=\{\cB:\ \pmb{\t}(\cB^n)=\pmb{\t}(\cA^n)\ for\ all\
 n\in\N\}.
\]
Another interesting equation $\pmb{\pi}(e^{\cA})=e^{\pmb{\t}(\cA)}$
for all $\cA\in\ol{\mH}$ immediately follows from \er{020}. Note
also that the resolvent \er{015} allows us to write closed form
expressions in the functional calculus, e.g.
$$
 f(\cA)=\frac1{2\pi i}\oint_{\pa\O}f(\l)(\l\cI-\cA)^{-1}d\l,\ \
 \pmb{\t}(f(\cA))=\frac1{2\pi
 i}\oint_{\pa\O}f(\l)\pmb{\t}((\l\cI-\cA)^{-1})d\l
$$
for analytic functions $f$ defined in some domain
$\O\supset\s(\cA)$. Let us finish with some exercises: {\it try to
extend the results of Theorem \ref{T3} to the arbitrary subset
$\a\ss\{0,...,N\}$, i.e. if $0\in\a$ then
$
 {\rm Inv}(\sum_{j\in\a}\mH_{j})=\prod_{j\in\a}\mF_j;
$ try to prove that $
 \prod_{r=j}^N\mF_r\lhd{\rm Inv}(\mH)
$ is a normal subgroup and $\sum_{r=j}^N\mH_r\ss\mH$ is a two-sided
ideal for all $j$.} Let us specify the structure of the paper:
Section \ref{S1} contains the proofs of all theorems; Section
\ref{example} provides a simple example of applications of our
results to some concrete operator.

\section{Proof of Theorems \ref{T1}-\ref{T5}}\lb{S1}

There are a lot of different ways to prove these theorems. We try to
make the proof to be more or less elementary, except the last part
where we discuss the trace norm. In the last part we intensively use
properties of direct integrals \cite{RS} and determinants and traces
\cite{GGK} of compact operators. Note that the first four Lemmas
repeat the arguments from \cite{K3}. We present their Proofs in a
short form.

\begin{lemma}\lb{L1}
Suppose that two operators of the form \er{002} are equal
\[\lb{201}
 {\bf A}_0\cdot+{\bf
 A}_1\langle{\bf B}_1\cdot\rangle_1+...+{\bf A}_N\langle{\bf
 B}_N\cdot\rangle_N=\wt{\bf A}_0\cdot+\wt{\bf
 A}_1\langle\wt{\bf B}_1\cdot\rangle_1+...+\wt{\bf A}_N\langle\wt{\bf
 B}_N\cdot\rangle_N.
\]
Then its components are equal too
\[\lb{202}
 {\bf A}_0=\wt{\bf A}_0\ \ and\ \ {\bf A}_j\langle{\bf
 B}_j\cdot\rangle_j=\wt{\bf A}_j\langle\wt{\bf
 B}_j\cdot\rangle_j\ \ for\ \ j=1,...,N.
\]
\end{lemma}
{\it Proof.} Suppose that ${\bf A}\ne{\bf A}_0$. Then there exists
some continuous vector-valued function ${\bf f}$ and ${\bf
k}^0\in[0,1]^N$ such that $({\bf A}_0-\wt{\bf A}_0){\bf f}({\bf
k}^0)={\bf f}^0\ne{\bf 0}$. Consider some continuous scalar function
$\chi({\bf k})$ with properties
\[\lb{203}
 \chi({\bf k})\le1,\ \ \chi({\bf k}^0)=1,\ \ \chi({\bf k})=0\ \ for\
 \ \|{\bf k}-{\bf k}^0\|>\ve.
\]
Then the identity \er{201} along with the continuity of ${\bf A}$,
${\bf B}$ and $\wt{\bf A}$, $\wt{\bf B}$ leads to
\[\lb{204}
 {\bf f}^0=({\bf A}_0-\wt{\bf A}_0)(\chi{\bf f})({\bf k}^0)=\sum_{j=1}^N({\bf A}_j\langle{\bf
 B}_j\chi{\bf f}\rangle_j-\wt{\bf A}_j\langle\wt{\bf
 B}_j\chi{\bf f}\rangle_j)({\bf k}^0).
\]
The fact that $|\langle\chi\rangle_j|\le2\ve$ shows that the norm of
the right-hand side of \er{204} is less than $C\ve$ with some fixed
$C$ depending on ${\bf A}$, ${\bf B}$ and $\wt{\bf A}$, $\wt{\bf B}$
only. This is the contradiction to a fixed norm of the left-hand
side of \er{204}. Thus ${\bf A}_0=\wt{\bf A}_0$. Now suppose that we
proved \er{202} up to $r-1$-th component for some $r\ge1$. So we
have the equality
\[\lb{205}
 {\bf
 A}_r\langle{\bf B}_r\cdot\rangle_r+...+{\bf A}_N\langle{\bf
 B}_N\cdot\rangle_N=\wt{\bf
 A}_r\langle\wt{\bf B}_r\cdot\rangle_r+...+\wt{\bf A}_N\langle\wt{\bf
 B}_N\cdot\rangle_N.
\]
Suppose that ${\bf A}_r\langle{\bf B}_r\cdot\rangle_r\ne\wt{\bf
 A}_r\langle\wt{\bf B}_r\cdot\rangle_r$. Then there exists
some continuous vector-value function ${\bf f}$ and ${\bf k}^0$ such
that
\[\lb{206}
 ({\bf A}_r\langle{\bf B}_r{\bf f}\rangle_r-\wt{\bf
 A}_r\langle\wt{\bf B}_r{\bf f}\rangle_r)({\bf k}^0)={\bf f}^0\ne{\bf
 0}.
\]
Let ${\bf k}^0_r=(k^0_{r+1},...,k^0_N)$ be the vector consisting of
$N-r$ components of the vector ${\bf k}^0$. Consider some continuous
scalar function $\chi({\bf k})$ with properties
\[\lb{207}
 \chi({\bf k})\le1,\ \ \chi([0,1]^r\ts{\bf k}^0_r)=1,\ \ \chi({\bf k})=0\ \ for\
 \ \|{\bf k}_r-{\bf k}^0_r\|>\ve,
\]
where ${\bf k}_r=(k_{r+1},...,k_N)$ is the vector consisting of
$N-r$ components of the vector ${\bf k}$. The identities \er{205},
\er{206} along with the continuity of ${\bf A}$, ${\bf B}$ and
$\wt{\bf A}$, $\wt{\bf B}$ lead to
\[\lb{208}
 {\bf f}^0=({\bf A}_r\langle{\bf B}_r\chi{\bf f}\rangle_r-\wt{\bf
 A}_r\langle\wt{\bf B}_r\chi{\bf f}\rangle_r)({\bf k}^0)=\sum_{j=r+1}^N({\bf A}_j\langle{\bf
 B}_j\chi{\bf f}\rangle_j-\wt{\bf A}_j\langle\wt{\bf
 B}_j\chi{\bf f}\rangle_j)({\bf k}^0).
\]
The fact that $|\langle\chi\rangle_j|\le2\ve$ for $j\ge r+1$ shows
that the norm of the right-hand side of \er{208} is less than $C\ve$
with some fixed $C$ depending on ${\bf A}$, ${\bf B}$ and $\wt{\bf
A}$, $\wt{\bf B}$ only. This is the contradiction to a fixed norm of
the left-hand side of \er{208}. Thus ${\bf A}_r\langle{\bf
B}_r\cdot\rangle_r=\wt{\bf A}_r\langle\wt{\bf B}_r\cdot\rangle_r$
and we finish proof by induction. \BBox

\begin{lemma}\lb{L2}
Consider the operator $\cA$ of the form \er{002}. Suppose that
$\det{\bf A}_0({\bf k}^0)=0$ at some ${\bf k}^0\in[0,1]^N$. Then
$\cA$ is non-invertible.
\end{lemma}
{\it Proof.} Let ${\bf f}^0$ be the corresponding null-vector ${\bf
A}_0({\bf k}^0){\bf f}^0={\bf 0}$ with Hilbert norm $\|{\bf
f}^0\|=1$. Without loss of generality we may assume that ${\bf
k}^0\in(0,1)^N$. For all sufficiently small $\ve>0$ define scalar
functions
\[\lb{209}
 \chi_{\ve}({\bf k})=\ca \ve^{-\frac N2}, & \max\limits_{j=1,...,N}|k_j-k_j^0|<\ve/2,\\
                   0, & otherwise. \ac
\]
The Hilbert norm of functions ${\bf f}_{\ve}({\bf
k})=\chi_{\ve}({\bf k}){\bf f}_0$ is equal to $1$ but the norm of
\[\lb{210}
 \cA{\bf f}_{\ve}={\bf A}_0{\bf f}_{\ve}+\sum_{j=1}^N{\bf A}_j\langle{\bf
 B}_j\chi{\bf f}_{\ve}\rangle_j
\]
tends to $0$ for $\ve\to0$ because the support of ${\bf f}_{\ve}$
tends to ${\bf k}_0$, ${\bf A}_0({\bf k}^0){\bf f}^0={\bf 0}$, all
matrix-valued functions ${\bf A}$, ${\bf B}$ are continuous and
Hilbert norm of $\langle\chi\rangle_j$ is equal to $\ve^{\frac j2}$
and tends to $0$ for $\ve\to0$. The Banach Theorem about continuous
inverse operators shows us that $\cA$ is non-invertible. \BBox

\begin{lemma}\lb{L3}
Consider the operator \er{002} of the special form
\[\lb{211}
 \cA=\cI+{\bf A}_r\langle{\bf
 B}_r\cdot\rangle_r+\sum_{j=r+1}^N{\bf A}_j\langle{\bf
 B}_j\cdot\rangle_j
\]
with some $r\ge1$. If
\[\lb{212}
 \det({\bf I}+\langle{\bf B}_r{\bf
 A}_r\rangle_r)({\bf k}^0)=0
\]
at some ${\bf k}^0\in[0,1]^{N-r}$ (the determinant does not depend
on the first $r$ components of ${\bf k}$) then $\cA$ is
non-invertible.
\end{lemma}
{\it Proof.} Let ${\bf f}^0$ be a null-vector of the matrix $({\bf
I}+\langle{\bf B}_r{\bf A}_r\rangle_r)({\bf k}^0)$ with the Hilbert
norm $\|{\bf f}^0\|=1$. Without loss of generality we may assume
that ${\bf k}^0\in(0,1)^N$. For all sufficiently small $\ve>0$
define scalar functions
\[\lb{213}
 \chi_{\ve}({\bf k})=\ca \ve^{-\frac {N-r}2}, & \max\limits_{j=r+1,...,N}|k_j-k_j^0|<\ve/2,\\
                   0, & otherwise \ac
\]
and vector-valued functions ${\bf f}_{\ve}=\chi_{\ve}{\bf A}_r{\bf
f}^{0}$. Suppose that the Hilbert norm of functions $\|{\bf
f}_{\ve}\|$ tends to zero for $\ve\to0$. Then the fact that
$\|\chi_{\ve}{\bf f}^0\|=1$ gives us
\[\lb{214}
 \|\chi_{\ve}{\bf f}^0+\langle\chi_{\ve}{\bf B}_r{\bf A}_r{\bf
 f}^0\rangle_r\|=\|\chi_{\ve}{\bf f}^0+\langle{\bf B}_r{\bf f}_{\ve}\rangle_r\|=1+o(1),\ \
 \ve\to0.
\]
On the other hand
\[\lb{215}
 \|\chi_{\ve}{\bf f}^0+\langle\chi_{\ve}{\bf B}_r{\bf A}_r{\bf
 f}^0\rangle_r\|=\|\chi_{\ve}({\bf I}+\langle{\bf B}_r{\bf A}_r\rangle_r){\bf f}_0\|=o(1),\ \
 \ve\to0
\]
because ${\bf f}_0$ is a null-vector of $({\bf I}+\langle{\bf
B}_r{\bf A}_r\rangle_r)({\bf k}^0)$, the function $\chi_{\ve}$ does
not depend on the first $r$ components of ${\bf k}$ and its support
tends to $[0,1]^{r}\ts{\bf k}^0$ for $\ve\to0$. The formulas
\er{214} and \er{215} contradict each other, which means that our
assumption is not clear and we have that
\[\lb{216}
 \|{\bf f}_{\ve}\|\not\to0\ \ for\ \ \ve\to0.
\]
The identity \er{211} and the definition of ${\bf f}_{\ve}$ give us
\[\lb{217}
 \cA{\bf f}_{\ve}=\chi_{\ve}{\bf A}_{r}({\bf I}+\langle{\bf B}_r{\bf A}_r\rangle_r){\bf
 f}^0+\sum_{j=r+1}^N{\bf A}_j\langle\chi_{\ve}{\bf
 B}_j{\bf A}_r{\bf f}^0\rangle_j,
\]
which leads to
\[\lb{218}
 \|\cA{\bf f}_{\ve}\|\to\ve\ \ for\ \ \ve\to0,
\]
since we have arguments \er{215}, continuity of ${\bf A}$, ${\bf B}$
and $\|\langle\chi_{\ve}\rangle_j\|=\ve^{\frac{j-r}2}$ tends 0 for
$\ve\to0$ and $j>r$. The formulas \er{216}, \er{218} and the Banach
Theorem about continuous inverse operators show us that $\cA$ is
non-invertible. \BBox

\begin{lemma}\lb{L4} The set $\mH_{j}$ \er{010} is an algebra.
\end{lemma}

{\it Proof.} It follows from the following identities:
\[\lb{219}
 {\bf A}_j\langle{\bf B}_j\cdot\rangle_{j}+\wt{\bf A}_j\langle\wt{\bf B}_j\cdot\rangle_{j}=
 \wt{\bf C}_j\langle\wt{\bf D}_j\cdot\rangle_{j},
\]
where
\[\lb{220}
 \wt{\bf C}_j=\ma {\bf A}_j & \wt{\bf A}_j \am,\ \ \wt{\bf D}_j=\ma {\bf B}_j \\ \wt{\bf B}_j
 \am
\]
and
\[\lb{221}
 \lt({\bf A}_j\langle{\bf B}_j\cdot\rangle_{j}\rt)\circ\lt(\wt{\bf A}_r\langle\wt{\bf B}_r\cdot\rangle_{r}\rt)=
 {\bf A}_j\langle{\bf B}_j\wt{\bf A}_r\langle\wt{\bf B}_r\cdot\rangle_{r}\rangle_{j}
 =\wt{\bf C}_s\langle\wt{\bf D}_s{\bf u}\rangle_{s},
\]
where
\[\lb{222}
 s=\max\{j,r\}\ \ and \ \ca \wt{\bf C}_s={\bf A}_j\langle{\bf B}_j\wt{\bf A}_r\rangle_{j},\ \ \wt{\bf D}_s=\wt{\bf B}_r,\ \ j\le r \\
 \wt{\bf C}_s={\bf A}_j,\ \ \wt{\bf D}_s=\langle{\bf B}_j\wt{\bf A}_r\rangle_{r}\wt{\bf B}_r, \ \  j>r
 \ac.\ \ \BBox
\]

\begin{lemma}\lb{L5} The set $\mF_j$ \er{012a} is a group for any $j=0,...,N$. If $j\ge1$ then
the element $\cA=\cI+{\bf A}\langle{\bf B}\cdot\rangle_j$ belongs to
$\mF_j$ if and only if the determinant of ${\bf E}={\bf
I}+\langle{\bf B}{\bf A}\rangle_j$ is non-zero everywhere. In this
case the inverse operator is
\[\lb{223}
 \cA^{-1}=\cI-{\bf A}{\bf E}^{-1}\langle{\bf B}\cdot\rangle_j.
\]
\end{lemma}
{\it Proof.} For $j=0$ the statement is trivial. Consider the case
$j\ge1$. If $\cA,\cB\in\mF_j$ then by \er{221}, \er{222} the element
$\cC=\cA\circ\cB$ has the form $\cC=\cI+{\bf C}\langle{\bf
D}\cdot\rangle_j$ with some continuous matrix-valued functions ${\bf
C}$, ${\bf D}$ and hence it belongs to $\mF_j$ because it is
invertible like $\cA$ and $\cB$.

Let $\cA=\cI+{\bf A}\langle{\bf B}\cdot\rangle_j$ be some element of
$\mF_j$. If $\det{\bf E}=0$ at some point then by Lemma \ref{L3} the
operator $\cA$ is non-invertible, which is impossible because
$\cA\in\mF_j$. Then $\det{\bf E}\ne0$ everywhere and hence ${\bf
E}^{-1}$ is a continuous matrix-valued function. Define
$\cB=\cI-{\bf A}{\bf E}^{-1}\langle{\bf B}\cdot\rangle_j$. Then
$$
 \cA\circ\cB=\cI+{\bf A}\langle{\bf B}\cdot\rangle_j-
 {\bf A}{\bf E}^{-1}\langle{\bf B}\cdot\rangle_j-{\bf A}\langle{\bf B}{\bf A}{\bf
 E}^{-1}\rangle_j\langle{\bf B}\cdot\rangle_j=
$$
$$
 \cI+({\bf A}-{\bf A}{\bf E}^{-1}-{\bf A}\langle{\bf B}{\bf A}\rangle{\bf E}^{-1})\langle{\bf
 B}\cdot\rangle_j=\cI+({\bf A}-{\bf A}{\bf E}{\bf E}^{-1})\langle{\bf
 B}\cdot\rangle_j=\cI,
$$
where we used the fact that ${\bf E}$ does not depend on the first
$j$ components of the vector ${\bf k}$. \BBox

{\bf Proof of Theorem \ref{TG}.} It follows from Lemmas \ref{L2} and
\ref{L5}. \BBox

{\bf Proof of Theorem \ref{T1}.}

\underline{Step 0.} If $\pi_0=\det{\bf A}_0({\bf k}^0)=0$ at some
point ${\bf k}^0\in[0,1]^N$ then by Lemma \ref{L2} the operator
$\cA$ is non-invertible. Suppose that $\det{\bf A}_0\ne0$
everywhere. Then ${\bf A}_0^{-1}$ is a continuous matrix-valued
function and we may define the operator (see \er{005})
\[\lb{224}
 \cA_0={\bf A}_0^{-1}\cA=\cI+{\bf A}_{10}\langle{\bf
 B}_1\cdot\rangle_{1}+...+{\bf A}_{N0}\langle{\bf
 B}_N\cdot\rangle_{N}.
\]

\underline{Step 1.} If $\pi_1=\det{\bf E}_1({\bf k}_1^0)=0$ at some
point ${\bf k}_1^0\in[0,1]^{N-1}$ then by Lemma \ref{L3}  the
operator $\cA_0$ and hence $\cA$ (see \er{224}) are non-invertible.
Suppose that $\det{\bf E}_1\ne0$ everywhere. Then ${\bf E}_1^{-1}$
is a continuous matrix-valued function and we may define the
operator (see \er{223} and \er{006})
\[\lb{225}
 \cA_1=(\cI+{\bf A}_{10}\langle{\bf
 B}_1\cdot\rangle_{1})^{-1}\circ\cA_0=(\cI-{\bf A}_{10}{\bf E}^{-1}_1\langle{\bf
 B}_1\cdot\rangle_{1})\circ\cA_0=
\]
\[\lb{226}
 \cI+{\bf A}_{21}\langle{\bf
 B}_2\cdot\rangle_{2}+...+{\bf A}_{N1}\langle{\bf
 B}_N\cdot\rangle_{N}.
\]

\underline{Step 2.} If $\pi_2=\det{\bf E}_2({\bf k}_1^0)=0$ at some
point ${\bf k}_2^0\in[0,1]^{N-2}$ then by Lemma \ref{L3}  the
operator $\cA_1$ and hence $\cA_0$ and $\cA$ (see \er{224}-\er{226})
are non-invertible. Suppose that $\det{\bf E}_2\ne0$ everywhere.
Then ${\bf E}_2^{-1}$ is a continuous matrix-valued function and we
may define the operator (see \er{223} and \er{007})
\[\lb{225a}
 \cA_2=(\cI+{\bf A}_{21}\langle{\bf
 B}_2\cdot\rangle_{2})^{-1}\circ\cA_1=(\cI-{\bf A}_{21}{\bf E}^{-1}_2\langle{\bf
 B}_2\cdot\rangle_{2})\circ\cA_1=
\]
\[\lb{226a}
 \cI+{\bf A}_{32}\langle{\bf
 B}_3\cdot\rangle_{3}+...+{\bf A}_{N2}\langle{\bf
 B}_N\cdot\rangle_{N}.
\]
Repeating this procedure up to the step N we finish the proof. Note
that we also obtain the identities \er{013} and \er{015}. \BBox

{\bf Proof of Theorem \ref{C1}.} It follows immediately from the
proof of Theorem \ref{T1} and from the identity for inverse
operators \er{223}. \BBox

\begin{definition}\lb{D1}
For $j=1,...,N$ and for any two continues matrix-valued functions
${\bf A}$ and ${\bf B}$ of sizes $M\ts{M}_1$ and $M_1\ts M$ ($M_1$
is any positive integer) defined on $[0,1]^N$ introduce the
following scalar function
\[\lb{227}
 \wt\pi_j({\bf A},{\bf B})=\det({\bf I}+\langle{\bf B}{\bf
 A}\rangle_j).
\]
\end{definition}

\begin{lemma} \lb{L6} Let $\cA_i=\cI+{\bf A}_i\langle{\bf
B}_i\cdot\rangle_j$, $i=1,2$ be two arbitrary operators of the form
\er{002}. Then there exist continuous matrix-valued functions ${\bf
A}_3$, ${\bf B}_3$ satisfying
\[\lb{228}
 \cA_1\circ\cA_2=\cI+{\bf A}_3\langle{\bf B}_3\cdot\rangle_j,\ \
 \wt\pi_j({\bf A}_1,{\bf B}_1)\wt\pi_j({\bf A}_2,{\bf
 B}_2)=\wt\pi_j({\bf A}_3,{\bf B}_3).
\]
\end{lemma}
{\it Proof.} Consider the composition
$$
 \cA_1\circ\cA_2=(\cI+{\bf A}_1\langle{\bf
 B}_1\cdot\rangle_j)\circ(\cI+{\bf A}_2\langle{\bf
B}_2\cdot\rangle_j)=
$$
$$
\cI+{\bf A}_1\langle{\bf
 B}_1\cdot\rangle_j+{\bf A}_2\langle{\bf
 B}_2\cdot\rangle_j+{\bf A}_1\langle{\bf
 B}_1{\bf A}_2\rangle_j\langle{\bf
 B}_2\cdot\rangle_j=\cI+{\bf A}_3\langle{\bf B}_3\cdot\rangle_j
$$
with
$$
 {\bf A}_3=\ma {\bf A}_1 & {\bf A}_2+{\bf A}_1\langle{\bf
 B}_1{\bf A}_2\rangle_j\am,\ \ {\bf B}_3=\ma {\bf B}_1 \\ {\bf B}_2
 \am.
$$
Then
$$
 \wt\pi_j({\bf A}_3,{\bf B}_3)=\det({\bf I}+\langle{\bf B}_3{\bf
 A}_3\rangle_j)=
$$
$$
 \det\ma {\bf I}+\langle{\bf B}_1{\bf A}_1\rangle_j &
 \langle{\bf B}_1{\bf A}_2\rangle_j+\langle{\bf B}_1{\bf A}_1\rangle_j\langle{\bf B}_1{\bf A}_2\rangle_j \\
 \langle{\bf B}_2{\bf A}_1\rangle_j & {\bf I}+\langle{\bf B}_2{\bf
 A}_2\rangle_j+\langle{\bf B}_2{\bf A}_1\rangle_j\langle{\bf B}_1{\bf A}_2\rangle_j
 \am=
$$
$$
 \det\ma {\bf I}+\langle{\bf B}_1{\bf A}_1\rangle_j & {\bf 0} \\ {\bf 0} & {\bf I} \am
 \det\ma {\bf I} & {\bf 0} \\ \langle{\bf B}_2{\bf A}_1\rangle_j & {\bf I}\am
 \det\ma {\bf I} & \langle{\bf B}_1{\bf A}_2\rangle_j \\ {\bf 0} & {\bf I}+\langle{\bf B}_2{\bf
 A}_2\rangle_j\am=
$$
$$
 \det({\bf I}+\langle{\bf B}_1{\bf A}_1\rangle_j)\det({\bf I}+\langle{\bf B}_2{\bf
 A}_2\rangle_j)=\wt\pi_j({\bf A}_1,{\bf B}_1)\wt\pi_j({\bf A}_2,{\bf
 B}_2).\ \ \BBox
$$

\begin{lemma}\lb{L7} Suppose that $\cA=\cI+{\bf A}\langle{\bf
B}\cdot\rangle_j=\cI$ is an identity operator. Then $\wt\pi_j({\bf
A},{\bf B})=1$.
\end{lemma}
{\it Proof.} Acting $\cA$ on each column of the matrix ${\bf A}$ and
after that multiplying by ${\bf B}$ and integrating we deduce that
$$
 {\bf A}\langle{\bf B}{\bf A}\rangle_j={\bf 0}\ \ \Rightarrow\ \ \langle{\bf B}{\bf A}\rangle_j^2={\bf
 0}.
$$
Then
$$
 1=\det{\bf I}=\det({\bf I}-t^2\langle{\bf B}{\bf A}\rangle_j^2)=\det({\bf I}+t\langle{\bf B}{\bf
 A}\rangle_j)\det({\bf I}-t\langle{\bf B}{\bf
 A}\rangle_j)=f(t)f(-t),
$$
where $t\in\C$ and $f(t)=\det({\bf I}+t\langle{\bf B}{\bf
A}\rangle_j)$ is a polynomial in $t$. Then $f(t)$ is a constant and
$f(t)=f(0)=1$. At the same time $\wt\pi_j({\bf A},{\bf B})=f(1)=1$.
\BBox

\begin{lemma}\lb{L8}
The following implication is fulfilled
\[\lb{229}
 \cI+{\bf A}_1\langle{\bf B}_1\cdot\rangle_j=\cI+{\bf A}_2\langle{\bf
 B}_2\cdot\rangle_j\in\mF_j\ \ \Rightarrow\ \ \wt\pi_j({\bf A}_1,{\bf B}_1)=\wt\pi_j({\bf A}_2,{\bf
 B}_2).
\]
\end{lemma}
{\it Proof.} Taking the inverse operator (see \er{223} in Lemma
\ref{L5}) and using \er{229} we have two identities
\[\lb{230}
 (\cI+{\bf A}_1\langle{\bf B}_1\cdot\rangle_j)\circ(\cI-{\bf A}_1{\bf E}_1^{-1}\langle{\bf
 B}_1\cdot\rangle_j)=\cI,
\]
\[\lb{231}
 (\cI-{\bf A}_1{\bf E}_1^{-1}\langle{\bf
 B}_1\cdot\rangle_j)\circ(\cI+{\bf A}_2\langle{\bf
 B}_2\cdot\rangle_j)=\cI,
\]
where ${\bf E}_1={\bf I}+\langle{\bf B}_1{\bf A}_1\rangle_j$. Then
Lemmas \ref{L6} and \ref{L7} give us
\[\lb{232}
 \wt\pi_j({\bf A}_1,{\bf B}_1)\wt\pi_j(-{\bf A}_1{\bf E}_1^{-1},{\bf
 B}_1)=1=\wt\pi_j(-{\bf A}_1{\bf E}_1^{-1},{\bf
 B}_1)\wt\pi_j({\bf A}_2,{\bf B}_2),
\]
which leads to $\wt\pi_j({\bf A}_1,{\bf B}_1)=\wt\pi_j({\bf
A}_2,{\bf B}_2)$. \BBox

\begin{definition}\lb{D2}
For any $j=1,...,N$ and any $\cA\in\mF_j$ define the mapping
$\wt\pi_j(\cA)=\wt\pi_j({\bf A},{\bf B})$, where $\cA=\cI+{\bf
A}\langle{\bf B}\cdot\rangle_j$ is some representation of $\cA$. By
Lemma \ref{L8} this definition of $\wt\pi_j(\cA)$ is correct. Also
define $\wt\pi_0({\bf A}\cdot)=\det{\bf A}$ for any ${\bf
A}\cdot\in\mF_0$.
\end{definition}

\begin{lemma}\lb{L9}
The mapping $\wt\pi_j:\mF_j\to \mC_j$ given by the definition
\ref{D2} is a group homomorphism (see also definition of $\mC_j$
after \er{015}).
\end{lemma}
{\it Proof.} Now this result follows from Lemma \ref{L6}. \BBox

\begin{lemma}\lb{L10} Suppose that $\cA=\cA_j\cA_r$ for some
$\cA_j\in\mF_j$ and $\cA_r\in\mF_r$ and $j\ne r$. Then there exists
unique representation $\cA=\wt\cA_r\wt\cA_j$ with $\wt\cA_j\in\mF_j$
and $\wt\cA_r\in\mF_r$. The identities
$\wt\pi_j(\cA_j)=\wt\pi_j(\wt\cA_j)$ and
$\wt\pi_r(\cA_r)=\wt\pi_r(\wt\cA_r)$ are fulfilled. Moreover, if
$j<r$ then $\wt\cA_j=\cA_j$, if $r<j$ then $\wt\cA_r=\cA_r$.
\end{lemma}
{\it Proof.} Consider the case $1\le j < r$ (other cases can be
proved similarly). Take some representations of $\cA_j$ and $\cA_r$
$$
 \cA_j=\cI+{\bf A}_j\langle{\bf B}_j\cdot\rangle_j,\ \ \cA_r=\cI+{\bf A}_r\langle{\bf
 B}_r\cdot\rangle_r.
$$
Then the following identities are fulfilled
$$
 \cA=\cA_j\circ\cA_r=(\cI+{\bf A}_j\langle{\bf
 B}_j\cdot\rangle_j)\circ(\cI+{\bf A}_r\langle{\bf
 B}_r\cdot\rangle_r)=
$$
$$
 \cI+{\bf A}_j\langle{\bf
 B}_j\cdot\rangle_j+{\bf A}_r\langle{\bf
 B}_r\cdot\rangle_r+{\bf A}_j\langle{\bf B}_j{\bf A}_r\rangle_j\langle{\bf
 B}_r\cdot\rangle_r=
$$
$$
 (\cI+({\bf A}_r+{\bf A}_j\langle{\bf B}_j{\bf A}_r\rangle_j)\langle{\bf
 B}_r\cdot\rangle_r\circ(\cI+{\bf A}_j\langle{\bf
 B}_j\cdot\rangle_j)^{-1})\circ(\cI+{\bf A}_j\langle{\bf
 B}_j\cdot\rangle_j)=
$$
$$
 (\cI+({\bf A}_r+{\bf A}_j\langle{\bf B}_j{\bf A}_r\rangle_j)\langle{\bf
 B}_r\cdot\rangle_r\circ(\cI-{\bf A}_j{\bf E}_j^{-1}\langle{\bf
 B}_j\cdot\rangle_j))\circ(\cI+{\bf A}_j\langle{\bf
 B}_j\cdot\rangle_j)=
$$
$$
 (\cI+({\bf A}_r+{\bf A}_j\langle{\bf B}_j{\bf A}_r\rangle_j)(\langle{\bf
 B}_r\cdot\rangle_r-\langle{\bf B}_r{\bf A}_j{\bf E}_j^{-1}\langle{\bf
 B}_j\cdot\rangle_j\rangle_r))\circ\cA_j=
$$
$$
 (\cI+({\bf A}_r+{\bf A}_j\langle{\bf B}_j{\bf
 A}_r\rangle_j)\langle({\bf B}_r-\langle{\bf B}_r{\bf A}_j\rangle_j{\bf E}_j^{-1}{\bf
 B}_j)\cdot\rangle_r)\circ\cA_j=
$$
$$
 \wt\cA_r\circ\cA_j=\wt\cA_r\circ\wt\cA_j,
$$
where ${\bf E}_j={\bf I}+\langle{\bf B}_j{\bf A}_j\rangle_j$ (see
Lemma \ref{L5}), $\wt\cA_j=\cA_j$ and $\wt\cA_r=\cI+\wt{\bf
A}_r\langle\wt{\bf B}_r\cdot\rangle$ with
$$
 \wt{\bf A}_r={\bf A}_r+{\bf A}_j\langle{\bf B}_j{\bf
 A}_r\rangle_j,\ \ \wt{\bf B}_r={\bf B}_r-\langle{\bf B}_r{\bf A}_j\rangle_j{\bf E}_j^{-1}{\bf
 B}_j.
$$
Thus, we have $\wt\pi_j(\cA_j)=\wt\pi_j(\wt\cA_j)$ and
$$
 \wt\pi_r(\wt\cA_r)=\det({\bf I}+\langle\wt{\bf B}_r\wt{\bf
 A}_r\rangle_r)=\det({\bf I}+\langle{\bf B}_r{\bf A}_r\rangle_r+
$$
$$
 \langle{\bf B}_r{\bf A}_j\langle{\bf B}_j{\bf A}_r\rangle_j-
 \langle{\bf B}_r{\bf A}_j\rangle_j{\bf E}_j^{-1}{\bf B}_j{\bf A}_r-\langle{\bf B}_r{\bf A}_j\rangle_j{\bf E}_j^{-1}{\bf B}_j{\bf A}_j
 \langle{\bf B}_j{\bf A}_r\rangle_j\rangle_r)=\det({\bf I}+\langle{\bf B}_r{\bf A}_r\rangle_r+
$$
$$
 \langle\langle{\bf B}_r{\bf A}_j\rangle_j\langle{\bf B}_j{\bf A}_r\rangle_j-
 \langle{\bf B}_r{\bf A}_j\rangle_j{\bf E}_j^{-1}\langle{\bf B}_j{\bf A}_r\rangle_j-
 \langle{\bf B}_r{\bf A}_j\rangle_j{\bf E}_j^{-1}\langle{\bf B}_j{\bf
 A}_j\rangle_j \langle{\bf B}_j{\bf A}_r\rangle_j\rangle_r)=
$$
$$
 \det({\bf I}+\langle{\bf B}_r{\bf A}_r\rangle_r+\langle\langle{\bf B}_r{\bf A}_j\rangle_j({\bf I}-
 {\bf E}_j^{-1}-
 {\bf E}_j^{-1}\langle{\bf B}_j{\bf
 A}_j\rangle_j) \langle{\bf B}_j{\bf A}_r\rangle_j\rangle_r)=
$$
$$
 \det({\bf I}+\langle{\bf B}_r{\bf A}_r\rangle_r+\langle\langle{\bf B}_r{\bf A}_j\rangle_j({\bf I}-
 {\bf E}_j^{-1}({\bf I}+
 \langle{\bf B}_j{\bf
 A}_j\rangle_j)) \langle{\bf B}_j{\bf A}_r\rangle_j\rangle_r)=
$$
$$
 \det({\bf I}+\langle{\bf B}_r{\bf A}_r\rangle_r)=\wt\pi_r(\cA_r).
$$
Suppose that we have two different representations
$\cA=\wt\cA_r\wt\cA_j=\wh\cA_r\wh\cA_j$ with
$\wt\cA_r,\wh\cA_r\in\mF_r$ and $\wt\cA_j,\wh\cA_j\in\mF_j$. Then
$\mF_r\ni\wh\cA_r^{-1}\cA_r=\wh\cA_j\wt\cA_j^{-1}\in\mF_j$, which
gives us $\wt\cA_r=\wh\cA_r$ and $\wt\cA_j=\wh\cA_j$ because by
Lemma \ref{L1} we have that $\mF_r\cap\mF_j=\{\cI\}$ for $r\ne j$.
\BBox

\begin{lemma}\lb{L11}
The set $\mG$ defined in Theorem \ref{T2} is a group. For any
$\cA\in\mG$ there exists unique representation
\[\lb{233}
 \cA=\cA_0\circ\cA_1\circ...\circ\cA_N\ \ with\ \ \cA_j\in\mF_j.
\]
The mapping $\pmb{\pi}$ defined in \er{005a}-\er{008a} and \er{009}
has the form
\[\lb{234}
 \pmb{\pi}(\cA)=(\wt\pi_0(\cA_0),\wt\pi_1(\cA_1),...,\wt\pi_N(\cA_N)).
\]
\end{lemma}
{\it Proof.} If $\cA,\cB\in\mG$ then $\cA\circ\cB$ is invertible and
by Lemma \ref{L4} it belongs to $\mG$. The decomposition \er{233}
follows from the steps of Theorem \ref{T1}, see also its Proof and
\er{013}. The formula \er{015} discussed in the Proof of Theorem
\ref{T1} along with Lemma \ref{L4} gives us that $\cA^{-1}\in\mG$
and hence $\mG$ is a group. Suppose that we have two decompositions
$$
 \cA=\cA_0\circ\cA_1\circ...\circ\cA_N=\wt\cA_0\circ\wt\cA_1\circ...\circ\wt\cA_N.
$$
Then using $\cA_j^{-1}\in\mF_j$ and \er{221}, \er{222} we obtain
$$
 \wt\cA_0^{-1}\cA_0=\wt\cA_1\circ...\circ\wt\cA_N\circ(\cA_1\circ...\circ\cA_N)^{-1}=\cI+\{integral\
 operators\},
$$
which gives us $\wt\cA_0^{-1}\cA_0=\cI$ by Lemma \ref{L1}. Repeating
these arguments we deduce that $\wt\cA_j=\cA_j$ for all $j$. The
identity \er{234} follows from the definition of $\pi_j$ given in
Theorem \ref{T1}, its Proof and Definitions \ref{D1} and \ref{D2}.
\BBox

{\bf Proof of Theorem \ref{T2}.} Let $\cA,\cB\in\mG$ be two
operators. Consider their decompositions \er{233}
$$
 \cA=\cA_0\circ\cA_1\circ...\circ\cA_N,\ \
 \cB=\cB_0\circ\cB_1\circ...\circ\cB_N,\ \ \cA_j,\cB_j\in\mF_j.
$$
By Lemma \ref{L10}  we can rearrange the terms in the product
$\cA\circ\cB$ to obtain
\[\lb{235}
 \cA\circ\cB=\cA_0\circ\cA_1\circ...\circ\cA_N\circ\cB_0\circ\cB_1\circ...\circ\cB_N=
 \wt\cA_0\circ\wt\cB_0\circ...\circ\wt\cA_N\circ\wt\cB_N
\]
with
\[\lb{236}
 \wt\cA_j,\wt\cB_j\in\mF_j\ \ and\ \ \wt\pi_j(\wt\cA_j)=\wt\pi_j(\cA_j),\ \
 \wt\pi_j(\wt\cB_j)=\wt\pi_j(\cB_j).
\]
Denoting $\cC_j=\wt\cA_j\circ\wt\cB_j$ we obtain the unique
representation for the product (see Lemma \ref{L11})
\[\lb{237}
 \cA\circ\cB=\cC_0\circ...\circ\cC_N.
\]
Using \er{236} along with Lemma \ref{L9} we deduce that
\[\lb{238}
 \wt\pi_j(\cC_j)=\wt\pi_j(\wt\cA_j)\wt\pi_j(\wt\cB_j)=\wt\pi_j(\cA_j)\wt\pi_j(\cB_j),
\]
which with \er{234} give us
\[\lb{239}
 \pmb{\pi}(\cA\circ\cB)=\pmb{\pi}(\cA)\pmb{\pi}(\cB).\ \ \BBox
\]

{\bf Proof of Theorem \ref{T3}.} In general, these results are
similar to the results of Lemma \ref{L11} and can be obtained in the
same manner. \BBox

{\bf Proof of Theorem \ref{T4}.} i) First note that
$\pmb{\pi}(\cI+\cO(t))=\pmb{\pi}(\cI)+{\bf O}(t)$ for $t\to0$, where
$\cO$ and ${\bf O}$ are standard $O$-notations for bounded operators
and vectors. Now for any operator $\cA$ of the form \er{002} we have
that
\[\lb{240}
 \cI+t\cA=(\cI+t{\bf A}_0\cdot)\circ(\cI+t{\bf A}_1\langle{\bf
 B}_1\cdot\rangle_1)\circ...\circ(\cI+t{\bf A}_N\langle{\bf
 B}_N\cdot\rangle_N)\circ(\cI+\cO(t^2)),
\]
which leads to
\[\lb{241}
 \pmb{\pi}(\cI+t\cA)=\pmb{\pi}(\cI+t{\bf A}_0\cdot)\pmb{\pi}(\cI+t{\bf A}_1\langle{\bf
 B}_1\cdot\rangle_1)...\pmb{\pi}(\cI+t{\bf A}_N\langle{\bf
 B}_N\cdot\rangle_N)\pmb{\pi}(\cI+\cO(t^2))=
\]
\[\lb{242}
 \lt(\det({\bf I}+t{\bf A}_0),1,...1\rt)\lt(1,\det({\bf I}+t\langle{\bf B}_1{\bf
 A}_1\rangle_1),1,...1\rt)...
\]
\[\lb{243}
 ...\lt(1,...1,\det({\bf I}+t\langle{\bf B}_N{\bf
 A}_N\rangle_N)\rt)\lt(\pmb{\pi}(\cI)+{\bf O}(t^2)\rt)=
\]
\[\lb{244}
 \pmb{\pi}(\cI)+t(\Tr{\bf A}_0,\langle\Tr{\bf B}_1{\bf A}_1\rangle_1,...,\langle\Tr{\bf B}_N{\bf
 A}_N\rangle_N)+{\bf O}(t^2),
\]
which give us \er{018}. Note that in \er{241}-\er{244} we use the
standard asymptotics of $\det$ and the fact that
$\pmb{\pi}(\cI)=(1,...,1)$. The identities
\[\lb{245}
 \pmb{\pi}(\cI+t\a\cA+t\b\cB)=\pmb{\pi}\lt((\cI+t\a\cA)\circ(\cI+t\b\cB)\circ(\cI+\cO(t^2))\rt)=
\]
\[\lb{246}
 \pmb{\pi}(\cI+t\a\cA)\pmb{\pi}(\cI+t\b\cB)(\pmb{\pi}(\cI)+{\bf
 O}(t^2))=\pmb{\pi}(\cI)+t\a\pmb{\t}(\cA)+t\b\pmb{\t}(\cB)+{\bf O}(t^2)
\]
lead to the first formula in \er{019}. The identities
\[\lb{247}
 \pmb{\pi}(\cI-t^2\cB\circ\cA-t^2\cA^2-t^2\cB^2)=\pmb{\pi}\lt((\cI+t\cA)\circ(\cI+t\cB)\circ(\cI-t\cA-t\cB)\circ(\cI+\cO(t^3))\rt)=
\]
\[\lb{248}
 \pmb{\pi}\lt((\cI+t\cB)\circ(\cI+t\cA)\circ(\cI-t\cA-t\cB)\circ(\cI+\cO(t^3))\rt)=
\]
\[\lb{249}
 \pmb{\pi}(\cI-t^2\cA\circ\cB-t^2\cA^2-t^2\cB^2+\cO(t^3))
\]
lead to
\[\lb{250}
 \pmb{\t}(\cA^2+\cB^2+\cB\circ\cA)=\pmb{\t}(\cA^2+\cB^2+\cA\circ\cB),
\]
which with the first identity gives us the second identity in
\er{019}. \BBox

\begin{lemma}\lb{L12} Consider an operator $\cA={\bf A}\langle{\bf B}\cdot\rangle_j:L^2\to
L^2$. Then the spectrum of $\cA$ consists of eigenvalues. All
non-zero eigenvalues of $\cA$ coincide with non-zero eigenvalues of
the matrix ${\bf C}:=\langle{\bf B}{\bf A}\rangle_j$. The algebraic
multiplicities of these eigenvalues are the same for $\cA$ and ${\bf
C}$.
\end{lemma}
{\it Proof.} Without loss of generality we assume $j\not=0,N$. The
direct integral representation
\[\lb{252}
 \cA=\int^{\os}_{{\bf k}_j\in[0,1]^{N-j}}\cA({\bf k}_j),\ \ {\bf
 k}_{j}=(k_{j+1},...,k_N).
\]
gives us that the spectrum $\s(\cA)$ consists of eigenvalues
$\l({\bf k}_j)$ of the finite rank operators
\[\lb{253}
 \cA({\bf k}_j)={\bf
 A}({\bf k}_{\ol{j}},{\bf k}_j)\langle{\bf B}({\bf k}_{\ol{j}},{\bf
 k}_j)\cdot\rangle_j,\ \ {\bf
 k}_{\ol{j}}=(k_1,...,k_j).
\]
Now, it is not difficult to verify the following statements
$$
\cA({\bf k}_j){\bf u}_0({\bf k}_{\ol{j}},{\bf k}_j)=\l({\bf
k}_j){\bf u}_0({\bf k}_{\ol{j}},{\bf k}_j)\ \Rightarrow\ \ca {\bf
C}({\bf k}_j){\wt {\bf u}}_0({\bf k}_j)=\l({\bf k}_j){\wt {\bf
u}}_0({\bf k}_j),\\  {\wt {\bf u}}_0({\bf k}_j)=\langle{\bf B}({\bf
k}_{\ol{j}},{\bf
 k}_j){\bf u}_0({\bf k}_{\ol{j}},{\bf k}_j)\rangle_j,\ac
$$
$$
 \cA({\bf k}_j){\bf u}_1({\bf k}_{\ol{j}},{\bf k}_j)=\l({\bf k}_j){\bf u}_1({\bf
k}_{\ol{j}},{\bf k}_j)+{\bf u}_0({\bf k}_{\ol{j}},{\bf k}_j)\
\Rightarrow\ \ca{\bf C}({\bf k}_j){\wt {\bf u}}_1({\bf k}_j)=\l({\bf
k}_j){\wt
{\bf u}}_1({\bf k}_j)+{\wt {\bf u}}_0({\bf k}_j),\\
{\wt {\bf u}}_1({\bf k}_j)=\langle{\bf B}({\bf k}_{\ol{j}},{\bf
 k}_j){\bf u}_1({\bf k}_{\ol{j}},{\bf k}_j)\rangle_j,\ac
$$
$$
 {\bf
C}({\bf k}_j){\wt {\bf u}}_0({\bf k}_j)=\l({\bf k}_j){\wt {\bf
u}}_0({\bf k}_j)\ \Rightarrow\ \ca\cA({\bf k}_j){\bf u}_0({\bf
k}_{\ol{j}},{\bf k}_j)=\l({\bf k}_j){\bf u}_0({\bf k}_{\ol{j}},{\bf
k}_j),\\ {\bf u}_0({\bf k}_j,{\bf k}_{\ol{j}})={\bf
 A}({\bf k}_{\ol{j}},{\bf k}_j){\wt {\bf u}}_0({\bf k}_j),\ac
$$
$$
 {\bf C}({\bf k}_j){\wt {\bf u}}_1({\bf k}_j)=\l({\bf
k}_j){\wt {\bf u}}_1({\bf k}_j)+{\wt {\bf u}}_0({\bf k}_j)\
\Rightarrow\ \ca\cA({\bf k}_j){\bf u}_1({\bf k}_{\ol{j}},{\bf
k}_j)=\l({\bf k}_j){\bf u}_1({\bf k}_{\ol{j}},{\bf k}_j)+{\bf
u}_0({\bf k}_{\ol{j}},{\bf k}_j),\\ {\bf u}_1({\bf k}_j,{\bf
k}_{\ol{j}})={\bf
 A}({\bf k}_{\ol{j}},{\bf k}_j){\wt {\bf u}}_1({\bf k}_j)\ac
$$
These statements show the one-to-one correspondence between
eigenvalues and eigenvectors (including adjoint eigenvectors which
belong to Jordan blocks) of $\cA({\bf k}_j)$ and ${\bf C}({\bf
k}_j)$. \BBox

{\bf Proof of Theorem \ref{T5}.} Due to Lemma \ref{L1} and to the
fact that each summand of $\cA\in\mH$ \er{002} is a direct integral
of finite rank operators (see \er{252},\er{253}) we may write the
following isomorphism of linear spaces
\[\lb{254}
 \mH\simeq\int^{\os}_{{\bf k}\in[0,1]^N}\mS_0d{\bf k}\os\int^{\os}_{{\bf k}_1\in[0,1]^{N-1}}\mS_1d{\bf
 k}_1\os...\os\mS_N,
\]
where $\mS_j$ is an algebra of finite rank operators acting on
$L^2_{j,M}$. Taking for each $\cR\in\mS_j$ the trace norm
$\|\cR\|_{TR}={\rm Tr}(\cR^*\cR)^{\frac12}$ (see \cite{GGK}, Theorem
5.1) we obtain the norm on the direct integral $\int^{\os}_{{\bf
k}_j\in[0,1]^{N-j}}\mS_jd{\bf
 k}_j$:
$$
 \|\int^{\os}_{{\bf k}_j\in[0,1]^{N-j}}\cR({\bf k}_j)d{\bf
 k}_j\|_{\rm tr}=\max_{{\bf k}_j\in[0,1]^{N-j}}\|\cR({\bf
 k}_j)\|_{TR}.
$$
The sum of these norms for all $j$ coincides with the norm
$\|\cdot\|_{\rm tr}$ \er{024} on $\mH$ (we also use \er{254} and
Lemma \ref{L12} which allows us to compute the trace norm
explicitly).

Consider operators $\cA,\cB\in\mH$ and $\cC=\cA\circ\cB\in\mH$. They
have unique representations
$$
 \cA=\sum_{j=0}^N\cA_N,\ \ \cB=\sum_{j=0}^N\cB_j,\ \ \cC=\sum_{j=0}^N\cC_j,\ \ \cA_j,\cB_j,\cC_j\in\int^{\os}_{{\bf
 k}_j\in[0,1]^{N-j}}\mS_j.
$$
The operators $\cC_j$ are of the form (see \er{221})
$$
 \cC_j=\cA_j\circ\cB_j+\sum_{r=0}^{j-1}(\cA_r\circ\cB_j+\cA_j\circ\cB_r).
$$
Denoting the standard operator norm of operators acting on some
Hilbert space as $\|\cdot\|$ and using the fact that the standard
operator norm is weaker than the trace norm and the fact that the
trace norm is sub-multiplicative (see \cite{GGK}, Theorem (5.1) and
Eq. (2.6) on p. 51) we obtain (we use also the fact that the norm of
direct integrals is a maximum of integrands)
$$
 \|\cC_j\|_{\rm tr}\le\|\cA_j\|_{\rm tr}\|\cB_j\|_{\rm tr}+\sum_{r=0}^{j-1}(\|\cA_r\cB_j\|_{\rm tr}+\|\cA_j\cB_r\|_{\rm
 tr})\le\|\cA_j\|_{\rm tr}\|\cB_j\|_{\rm tr}+\sum_{r=0}^{j-1}(\|\cA_r\|\|\cB_j\|_{\rm tr}+\|\cA_j\|_{\rm
 tr}\|\cB_r\|)
$$
$$
 \le\|\cA_j\|_{\rm tr}\|\cB_j\|_{\rm tr}+\sum_{r=0}^{j-1}(\|\cA_r\|_{\rm tr}\|\cB_j\|_{\rm tr}+\|\cA_j\|_{\rm
 tr}\|\cB_r\|_{\rm tr}),
$$
which lead to $\|\cA\circ\cB\|_{\rm tr}\le\|\cA\|_{\rm
tr}\|\cB\|_{\rm tr}$ because $\|\cA\circ\cB\|_{\rm tr}=\|\cC\|_{\rm
tr}=\sum_{j=0}^N\|\cC_j\|_{\rm tr}$. Due to Lemma \ref{L12} and
\cite{GGK}, Corollary 3.4 we also obtain that
$\|\pmb{\t}(\cA)\|_{\rm c}\le\|\cA\|_{\rm tr}$ and then
$\|\pmb{\t}\|=1$ since $\|\pmb{\t}(\cI)\|_{\rm c}=\|\cI\|_{\rm tr}$.
Using \er{017} and the first identity of \er{019} we obtain that
\[\lb{251}
 \frac{\pa\pmb{\pi}(\l\cI-\cA)}{\pa
 \l}=\pmb{\pi}(\l\cI-\cA)\pmb{\t}\lt((\l\cI-\cA)^{-1}\rt)=\pmb{\pi}(\l\cI-\cA)\sum_{n=0}^{\iy}\frac{\pmb{\t}(\cA^n)}{\l^{n+1}},
\]
which after integration by $\l$ becomes \er{020}. The continuity of
$\pmb{\pi}$ follows from the continuity of $\pmb{\t}$, \er{020} and
the identity
$$
 \|\pmb{\pi}(\cA+\cB)-\pmb{\pi}(\cA)\|_{\rm c}\le\|\pmb{\pi}(\cA)\|_{\rm c}\|\pmb{\pi}(\cI+\cA^{-1}\cB)-\pmb{\pi}(\cI)\|_{\rm c},
$$
which tends to $0$ for $\|\cB\|_{\rm tr}\to0$ because
$\|\cdot\|_{\rm tr}$ is a sub-multiplicative norm. \BBox

\section{Example}\lb{example}

In this section we apply our method to some synthetic example of
integral operator. Let $N=2$ and $M=1$. Consider the following
self-adjoin operator acting on $L^2_{1,2}$
\[\lb{300}
 \cA u=-\int_0^1udk_1-f\int_0^1
 fudk_1-\int_0^1\int_0^1udk_1dk_2,\ \ u\in L^2_{2,1},
\]
where $f$ is some real continuous scalar function with
$\int_0^1fdk_1=0$ (for convenience). Taking $\l\cI-\cA$, $\l\in\C$
and using notations \er{001} we have
\[\lb{301}
 \l\cI-\cA=\l\cdot+\langle\cdot\rangle_1+f\langle
 f\cdot\rangle_1+\langle\cdot\rangle_2.
\]
The spectrum of $\cA$ is
\[\lb{302}
 \s(\cA)=\{\l:\ \l\cI-\cA\ is\ non-invertible\}.
\]
Using our scheme \er{005a}-\er{spec} we will calculate this spectrum
explicitly and with the "degree" (essential or discrete). In our
case the matrices ${\bf A}$, ${\bf B}$ (some of them are scalars,
see \er{002}) are
\[\lb{303}
 {\bf A}_0=\l,\ \ {\bf B}_0=1,\ \ {\bf A}_1=\ma 1 & f \am,\ \ {\bf
 B}_1=\ma 1 \\ f \am,\ \ {\bf A}_2=1,\ \ {\bf B}_2=1.
\]
On the \underline{Step 0} of Theorem \ref{T1} we have
\[\lb{304}
 \pi_0=\l,\ \ {\bf E}_0=\l,\ \ {\bf A}_{10}=\l^{-1}\ma 1 & f \am,\ \
 {\bf A}_{20}=\l^{-1}.
\]
On the \underline{Step 1} of Theorem \ref{T1} we have
\[\lb{305}
 \pi_1=\frac{(\l+1)(\l+\langle f^2\rangle_1)}{\l^2},\ \ {\bf E}_1=\ma 1+\l^{-1} & 0 \\ 0 & 1+\l^{-1}\langle f^2\rangle_1 \am,\ \
 {\bf A}_{21}=\frac{1}{\l+1}.
\]
On the last \underline{Step 2} of Theorem \ref{T1} we have
\[\lb{306}
 \pi_2=\frac{\l+2}{\l+1},\ \ {\bf E}_2=\frac{\l+2}{\l+1}.
\]
Thus the vector-valued determinant \er{009} of our operator
$\l\cI-\cA$ is
\[\lb{det_example}
 \pmb{\pi}\lt(\l\cdot+\langle\cdot\rangle_1+f\langle
 f\cdot\rangle_1+\langle\cdot\rangle_2\rt)=\lt(\l,\frac{(\l+1)(\l+\langle
 f^2\rangle_1)}{\l^2},\frac{\l+2}{\l+1}\rt).
\]
Due to Theorem \ref{T1} the condition {\it $\l\cI-\cA$ is
non-invertible} follows from the presence of zeroes $\pi_j$
(components of our determinant). Thus, in our case the spectrum is
\[\lb{307}
 \s(\cA)=\{0\}\cup\{-1\}\cup\{\l:\
 \l=-\langle f^2\rangle_1\ for\ some\
 k_2\}\cup\{-2\}.
\]
The "degree" of spectral points can be calculated with the function
\er{mult}
\[\lb{308}
 D(\l)=\ca 0, & \l=0,\\
           1, & \l=-1\ or\ \l=-\langle f^2\rangle_1\ne0,\\
           2, & \l=-2\ne-\langle f^2\rangle_1,\\
           3, & otherwise.
       \ac
\]
In particular $\l=-2$ is an isolated eigenvalue of $\cA$ iff
$\langle f^2\rangle_1\ne2$ for all $k_2\in[0,1]$. The Floquet-Bloch
dispersion curves (see remark before Theorem \ref{T2}) are of the
form
\[\lb{309}
 \ca \l_0({\bf k})=0, & {\bf k}\in[0,1]^2,\\
     \l_{1a}(k_2)=-1,& k_2\in[0,1],\\
     \l_{1b}(k_2)=-\langle f^2\rangle_1 & k_2\in[0,1],\\
           \l_2=-2. & \
       \ac
\]
For all $\l\not\in\s(\cA)$ the resolvent has the form (see \er{015})
\[\lb{310}
 (\l\cI-\cA)^{-1}=\l^{-1}\lt(\cI-\frac{\langle\cdot\rangle_2}{\l+2}\rt)\circ\lt(\cI-\frac{\langle\cdot\rangle_1}{\l+1}-
 \frac{f\langle f\cdot\rangle_1}{\l+\langle f^2\rangle_1}\rt).
\]
Due to \er{018} the trace of $\cA$ is
\[\lb{311}
 \pmb{\t}\lt(\l\cdot+\langle\cdot\rangle_1+f\langle
 f\cdot\rangle_1+\langle\cdot\rangle_2\rt)=(\l,1+\langle
 f^2\rangle_1,1).
\]
Due to \er{022}-\er{024} the trace norm of $\cA$ is
\[\lb{312}
 \|\l\cI-\cA\|_{\rm tr}=|\l|+2+\max_{k_2}\langle f^2\rangle_1.
\]
Taking component-wise logarithm of \er{det_example} and using
\er{020} we obtain
\[\lb{traces}
 \pmb{\t}(\cA^{n})=(-1)^n(0,1+\langle f^2\rangle_1^n,2^n-1).
\]

\section*{Acknowledgements}
This work was supported by the RSF project
N\textsuperscript{\underline{o}}15-11-30007.

\bibliography{group_short}

\end{document}